\documentstyle [11pt,epsf] {article}
\title{\vskip-15mm Interpreting the Cosmic Ray Composition
\thanks{Invited paper, to appear in "Topics in Cosmic-Ray Astrophysics",
M.A.\ DuVernois ed., (Nova Science Publishers, New-York), in press, 1999.}}
\author{Luke O'C Drury \and Jean-Paul Meyer \and Donald C Ellison}
\def\be{\begin{equation}}
\def\ee{\end{equation}}
\def\fo{f_{\rm O}}
\def\fref{f_{\rm ref}}
\def\fvol{f_{\rm vol}}
\begin{document} 

\maketitle

\section{Introduction}

Detailed composition measurements can be a very powerful means of
tracing origins, a fact used regularly by forensic scientists and art
historians. One of the main motivating factors for making detailed
observations of cosmic rays was always the hope that a unique
compositional signature could be found which pointed unambiguously to
a particular source. This has proven much harder than expected, but we
have now reached a point where it appears possible to begin to
decipher the information contained in the compositional data; the key,
we have discovered, is to read the data not in isolation, but in the
%
%
context provided by our general astronomical knowledge and by recent
developments in shock acceleration theory (Meyer, Drury and Ellison,
1997, 1998; Ellison, Drury and Meyer, 1997). In our view (not, it is only
fair to warn the reader, yet universally accepted) the data show
clearly that the Galactic cosmic ray particles originate predominantly
from the gas and dust of the general interstellar medium.

\section{What is the Composition?}

Before attempting to interpret the data it is important to be clear
about what exactly we are discussing. The raw measurements are the
charge-resolved, and in some cases mass-resolved, differential energy
spectra of the cosmic ray nuclei above the Earth's atmosphere. For
instrumental and statistical reasons, good measurements with clean
separation of the various species are only easily made for mildly
relativistic nuclei. The measurements are affected by solar modulation
at energies below a few GeV per nucleon.  By correcting for these
solar system effects we infer the local Galactic cosmic ray spectra
which would be observed in the interstellar medium just outside the
heliosphere. However it is clear that these in turn have been
influenced, in varying degrees, by spallation nuclear reactions and
other interactions during propagation through the interstellar
medium. If we attempt to correct for these propagation effects we
finally arrive at inferred {\em source spectra}. The relative fluxes
of the various nuclear species at fixed energy per nucleon (equivalently,
at fixed speed or Lorentz factor) in these demodulated and
depropagated spectra constitute what is usually called the Galactic
Cosmic Ray Source (GCRS) composition.

Quite a number of assumptions have already gone in at this stage. The
heliospheric corrections are small above a few GeV per nucleon and
probably uncontroversial. However the propagation corrections are
clearly dependent on the propagation model used and often implicitly
assume that there are distinct acceleration and propagation
phases. Particularly with the current interest in so-called
``reacceleration'' models for propagation, it is not clear that such a
sharp separation is justified. It should also be noted that most of
the published data have been ``de-propagated'' using the simple, but
clearly unphysical, leaky-box model of Galactic cosmic ray
confinement.  By talking loosely about ``the GCRS composition''
without specifying precisely the energy per nucleon or rigidity at
which the measurements were made we are also implicitly assuming that
all the species have virtually identical spectra in energy per
nucleon, an assumption which is approximately true for the main
nuclear components in the range from $1\, \rm GeV$ to $1\, \rm TeV$
per nucleon (in fact, the data suggests that helium has a slightly
flatter spectrum than hydrogen), but is certainly not true of the
electrons\footnote{One often sees statements that the electron to
proton ratio in the cosmic rays is 1 to 30 or 1 to 100.  Any such
statement is, however, meaningless if one does not specify how the
comparison is made. The above applies to comparison at a fixed energy.
If, by contrast, one compares at fixed Lorentz gamma
factor the electrons are much more abundant than the protons (as
pointed out by W Kundt)!}. It is worth noting that measuring at fixed
kinetic energy per nucleon (which is the form traditionally used in
experimental work) or fixed momentum per charge (ie rigidity, which is
often used in theoretical work) give essentially equal relative
abundances for all the heavy nuclei, but different values for the
hydrogen abundance relative to the heavies.

\section{Nuclear or Atomic Physics?}

The obvious first thing to do is to compare the abundance pattern seen
in the GCRS to the standard solar system pattern of abundances, which
appears to characterise all undifferentiated bodies in the solar
system including the sun itself. If corrections are made for the decay
of long-lived radioactive nuclides, giving what is sometimes called
the primordial solar-system or proto-solar abundance pattern, this
appears to be close to the general local Galactic pattern of
abundances (in as much as this can be determined); thus it has usually
been taken as the base-line ``standard'' composition.  However there
is now increasing evidence that, both in the local interstellar medium
(ISM) and in the surfaces of young B stars, the abundances of the
heavy elements relative to hydrogen are systematically {\em lower}
than in the sun by factors of order 1.5 to 2 (Snow and Witt, 1996, and
references therein). In contrast, the solar system abundances {\em
are} apparently typical of those in the local F and G type stars
(Edvardsson et al, 1993; Andersson and Edvardsson, 1994).  Bearing all
this in mind we will continue to use the solar system abundances as
reference values because they provide a well-determined set of values
for all the elements which one might reasonably expect to be relevant
to the local ISM, especially as regards the relative abundances of the
heavy elements.

\begin{figure}
\epsfxsize=\hsize
\epsfclipon\epsfbox[20 65 455 305]{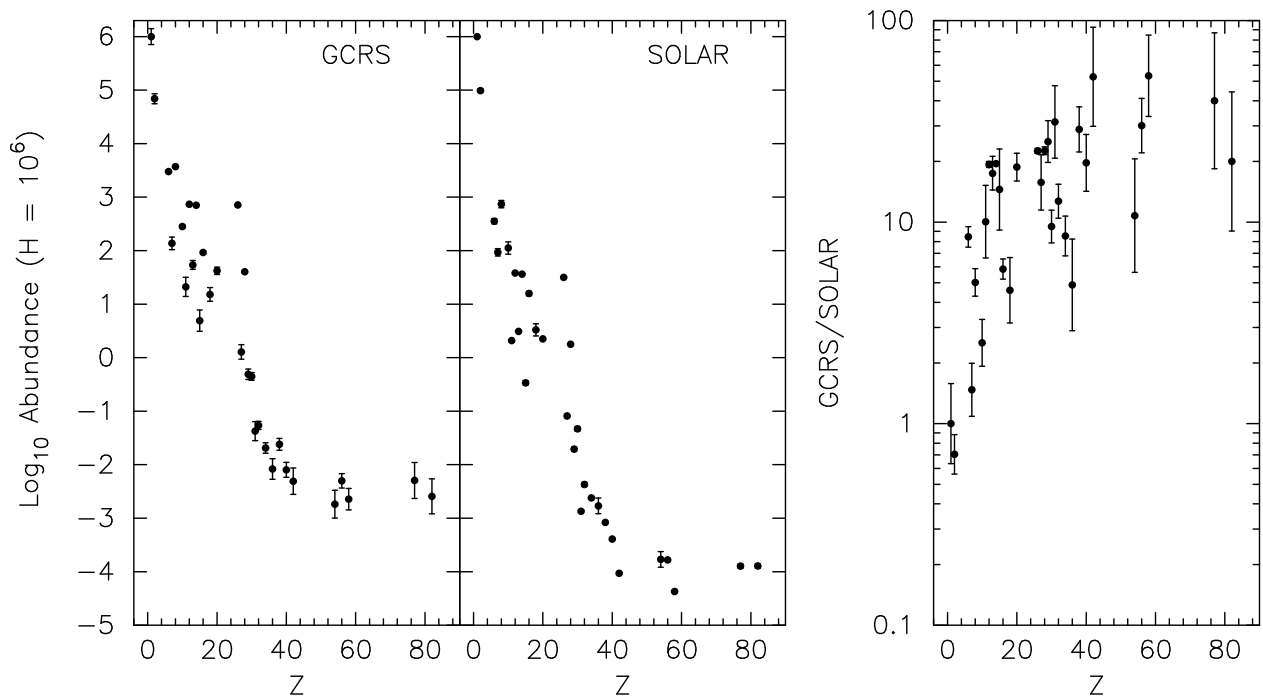}
\caption{\label{AbundancePlots}Plots of the GCRS and Solar abundances 
against atomic number for the major elements, both normalised to 
Hydrogen=$10^6$, and of the ratio GCRS/Solar.  A numerical table of 
these values can be found in Meyer, Drury and Ellison 1998}
\end{figure}

The GCRS and solar system abundances are compared in
Fig.~\ref{AbundancePlots} (see table in Meyer, Drury and Ellison 1998).  It
is immediately obvious that the GCRS composition is disappointingly
normal; all the elements are present, and the general pattern is
strikingly similar in both the GCRS and the solar system.  However
there are some significant differences, in particular Hydrogen is
deficient in the GCRS, or the heavy elements are enhanced, by quite
large factors relative to the solar system composition. For example
iron (Fe) and silicon (Si) are about a factor 30 higher relative to
hydrogen in the GCRS than in the solar system and this is a much
larger factor than any uncertainty in the measurements. Note that
adopting B~star abundances would make this even more extreme and
increase the overabundance of Fe and Si to between 45 and 60.  The
challenge is to interpret these slight (relative to the enormous
variations between the individual elements), but clearly significant,
differences between the two sets of abundances.

Now the heavy elements are known to be produced by nucleosynthesis in
stars and, for many elements specifically in supernova explosions, and
it has been suspected for a long time that cosmic rays are somehow
linked to the supernova phenomenon (this was first suggested in the
historic paper of Baade and Zwicky (1934) where they introduced the
name supernova, and cogently argued for on energetic grounds by
Ginzburg and Syrovatsky (1964) in their influential monograph). It is
therefore very natural to seek to interpret the differences between
the GCRS and local ISM (or solar system) abundances in terms of biases
stemming from the nuclear physics associated with nucleosynthesis,
perhaps during the slow core burning phase, but more likely during the
rapid explosive phase.  This effort was also stimulated by early
reports suggesting high GCRS abundances of the ultra-heavy
elements\footnote{The ultra-heavy elements are usually taken to be
those with nuclear charge greater than 28, although the term is used
very loosely}, including actinides, which are exclusively
produced during the supernova explosion.

It is now clear that these attempts to interpret the abundance
differences in terms of nucleosynthetic models are unconvincing. For
example, Ne is depleted by about a factor 8 relative to Mg, Al and Na
although all these elements are thought to be produced by C
burning. Similarly S and Ar are depleted by factors of order 4
relative to Si and Ca although these elements are all produced by O
and Si burning. No such large anomalies are found in supernova
nucleosynthesis calculations, especially for elements produced in the
same burning cycle (Woosley and Weaver, 1995; Timmes, Woosley and
Weaver, 1995; Arnett, 1995). By contrast, Mg, Al (C burning), Si and
Ca (O and Si burning) and Fe and Ni (e-process, i.e., explosive phase)
are present in the GCRS in proportions within 20\% of the solar
values. However Mg, Al, Si and Ca are synthesised in core-collapse
type II supernovae while Fe and Ni are predominantly produced in type
Ia supernovae and the nucleosynthesis calculations of different
supernova models typically yield deviations of these ratios by factors
of order 2.  In addition the ultra-heavy s-nuclei beyond $A=90$, which
are not produced in any type of supernova, are not underabundant
relative to the above elements, or to the r-nuclei, and the general
ultra-heavy abundances are not anomalously high.  Further, with the
exception of $^{22}$Ne/$^{20}$Ne (and possibly $^{13}$C/$^{12}$C and
$^{18}$O/$^{16}$O) all isotopic ratios are consistent with solar
values.

\begin{figure}
\epsfxsize=\hsize
\epsfbox{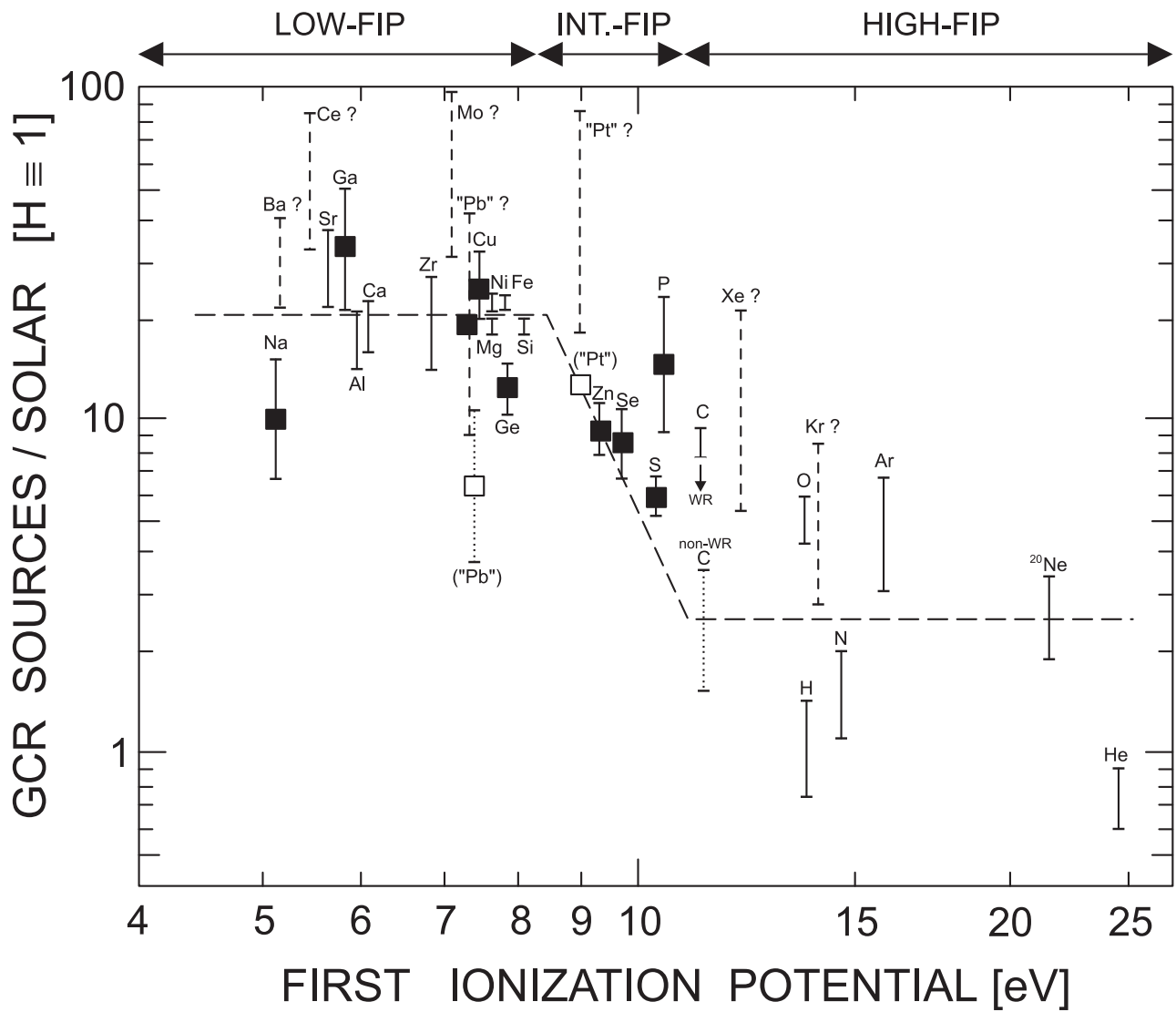}
\caption{\label{FIPPlot}The GCRS to Solar abundance ratios plotted
versus FIP. Solid squares denote those elements which can be used to
distinguish between FIP and volatility. For detailed discussion see
Meyer, Drury and Ellison 1998.}
\end{figure}

Remarkably, if the data are organised not by nuclear but by atomic 
properties some, though not {\em all}, of the differences can be 
accounted for.  In particular if the ratio of the GCRS abundance to 
the solar system abundance is plotted against the first ionization 
potential (FIP) of each element a definite pattern, the so-called FIP 
effect, is evident (see Fig.~\ref{FIPPlot}).  
There is a large group of low-FIP elements which show a roughly 
constant and large enhancement of about 30 relative to H. At a FIP of 
order $10\,\rm eV$ there is a rather sharp break and the elements with 
larger FIP values show much smaller, but more scattered, enhancements.


Now of course the first ionization potential measures how easy it is
to remove one electron from the outermost shell of electrons in the
ground state of the neutral atom, and thus correlates strongly with
chemistry. For example, elements with easily removed electrons tend to
be metallic and chemically very reactive forming stable compounds
which readily condense at high temperatures; elements with filled
outer shells have very firmly attached electrons and are the inert
gases which do not condense except under extreme laboratory
conditions.  So there exists, by and large, a relationship between the
FIP of the various elements and their volatility, which is
conveniently measured by the so-called condensation
temperature\footnote{The exact definition of the condensation
temperature is rather artificial.  One imagines starting with a sample
of solar composition gas at high temperature and a constant pressure
of $10^{-4}\,\rm atm$ and gradually lowering the temperature. The
condensation temperature is the temperature at which 50\% of the
dominant solid compound formed by each element has condensed out of
the gas phase.}.  It has long been known that the GCRS composition data
can also be organised in terms of this condensation temperature;
refractory (low-FIP) elements tend to be overabundant relative to
volatile (high-FIP) ones.  Fortunately, there are a few elements which
do not follow the general FIP/volatility correlation and which, in
principle, allow a distinction to be made between a FIP effect and a
volatility effect in the GCRS composition.  However these are not the
easiest of elements to measure!  Such data as is available tends to
favour volatility rather than FIP as the better organising parameter
(Meyer, Drury and Ellison 1997).  However it is clear that neither FIP
nor volatility alone completely accounts for the observations. In
particular a simple two-step volatility or FIP bias does {\em not}
account for the low relative abundances of H and He, the two most abundant
elements! There must be some additional effect, parameter or process
involved.

\begin{figure}
\epsfxsize=\hsize
\epsfbox{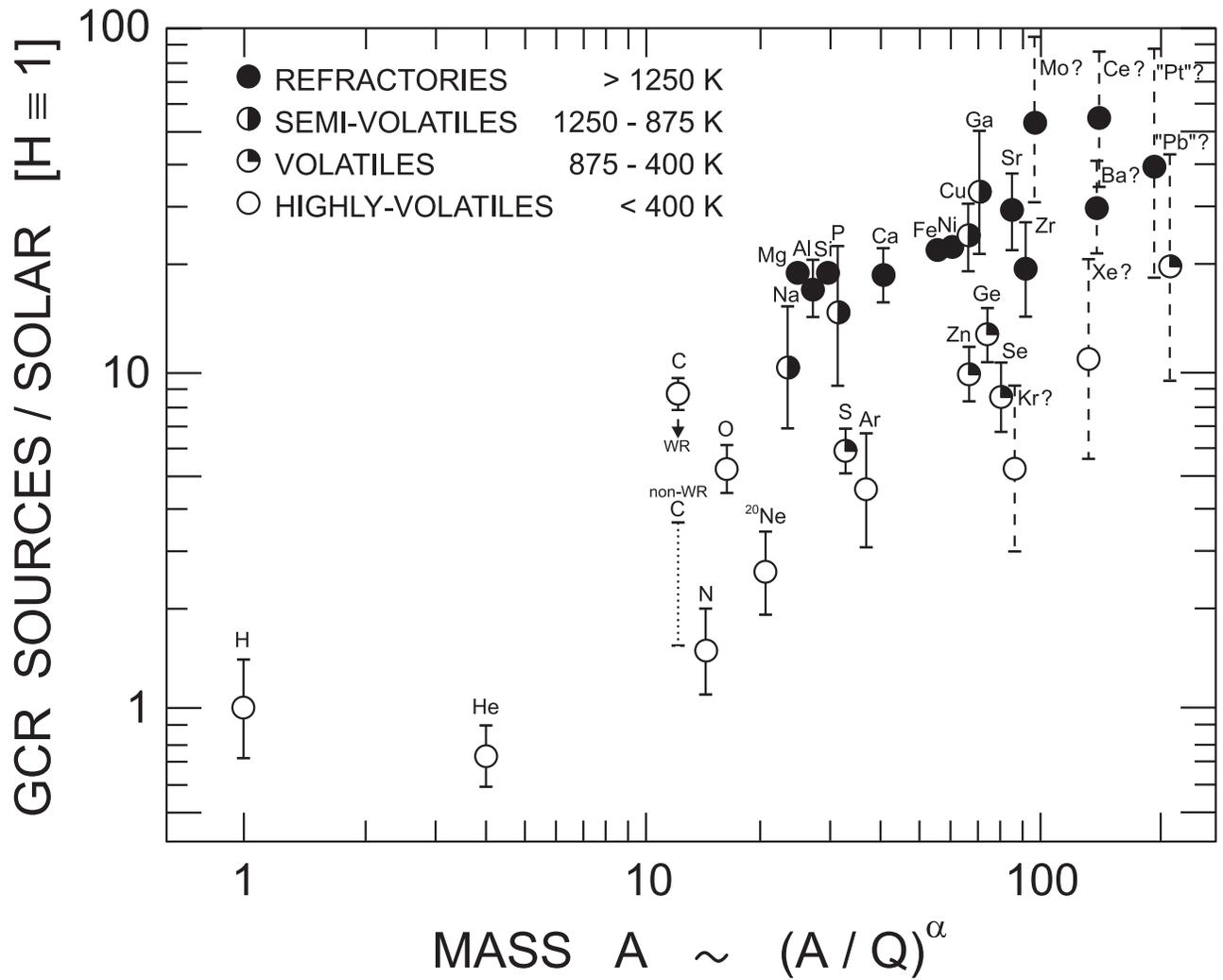}
\caption{\label{VolMassPlot}The GCRS to solar abundance ratios plotted
against element atomic mass number $A$ for four volatility classes.}
\end{figure}

In Fig.~\ref{VolMassPlot} we sort the elements according to their
volatility, and
then plot their abundance enhancements versus the element mass $A$.
This is a very interesting plot.  It first shows that the refractory
elements are globally enhanced relative to the volatile ones.  Among
the volatile elements the enhancements of all the inert gases and N
appear to follow a smoothly increasing function of the mass (roughly
$\propto A^{0.8}$); however, volatile H, C and O lie above this
correlation.  Among the refractories, by contrast, the enhancements
are roughly independent of the mass\footnote{It is important to note
that a tentative similar ordering of the data in terms of a combined
FIP and mass effect would {\em not} order the data as satisfactorily.
Specifically, the non-solar values of the GCRS abundance ratios
between elements of similar FIP {\em and} mass, but widely different
volatilities (Na/Mg, P/S, Ge/Fe, and Pb/Pt), cannot be interpreted in
terms of a combined FIP and mass fractionation (Meyer, Drury and
Ellison 1998).}.

In summary, the empirical evidence is that the GCRS composition is
basically similar to that of the local ISM and has not been affected
by specific nucleosynthetic processes (with the exception of the
$^{22}$Ne and associated $^{12}$C excesses). But it does show clear
signs of modification by factors depending on atomic physics or
chemistry (an enhancement of low-FIP or, more probably, refractory
elements) as well as on the element mass. It is very remarkable that
the composition of fully-stripped relativistic nuclei should show
traces of atomic physics effects with characteristic energies of only
a few electron volts, and this is clearly a significant clue to the
cosmic ray origin.

\section{The solar coronal FIP effect}

The FIP concept received a significant boost when it was discovered
that this effect operates in the atmosphere of the sun and biases the
composition of the solar corona, solar wind and solar energetic
particles. By some not entirely understood mechanism, ionized heavy
elements are preferentially lifted from the chromosphere into the
corona, giving a coronal composition enhanced relative to the bulk
solar composition in low FIP elements. This biased composition is then
also seen in the solar wind and in solar accelerated particles.

This remarkable discovery prompted attempts to relate the origin of
cosmic rays to the coronae of cool stars like the sun (Meyer,
1985). However it is certain that even if these stars are the source
of the accelerated material, they cannot be the source of the energy
needed for the acceleration. The only plausible known source of energy
capable of driving the acceleration processes remains the
mechanical explosion energy of supernovae. Thus this line of argument
requires the dwarf stars to somehow inject large amounts of FIP biased
but low energy ions (MeV) into the ISM which are later accelerated to the
observed energies by passing supernova shocks.

However there are problems in trying to make any such two-stage
model work {\em quantitatively} (Epstein, 1981). The basic problem is
that sub-relativistic ions suffer quite rapid ionization and Coulomb
energy losses in the ISM even allowing for the effect of electron
pick-up in screening the nuclear charge (Meyer, 1985). On the other
hand, the mean time interval between passages of strong supernova
shocks must be long in the general ISM (at least $10^7\,\rm y$);
otherwise the observed energy dependence of the secondary to primary
ratios could not be accounted for\footnote{These constraints would be
significantly alleviated if GCR production were located in regions of active
star formation containing many young low-mass stars with high levels
of surface activity and a few massive stars to provide supernovae, all
concentrated within a limited volume.}. In addition modern shock
acceleration
theory emphasises that the shock accelerates particles directly out of
the ``thermal'' distribution. Not only is there no need for a separate
pre-injected ``seed'' population: any such population, unless at
rather high number density\footnote{A rough estimate is that the
number density of MeV ions would need to be of order $10^{-6}$ that of
the background plasma, which in turn would imply an energy density in
the putative seed particles comparable to the thermal energy density
of the plasma.}, will tend to be swamped by the ISM particles
accelerated directly by the shock.

\section{SNR acceleration from the ISM}

As described elsewhere in this volume [x-refs here] much work has been
done on diffusive shock acceleration applied to supernova remnants
(SNRs) as a theoretical model for the origin of cosmic rays. However
this has mainly concentrated on the spectrum and the total power and,
until recently, the question of composition was largely ignored. It
has been known for a long time that shocks are intrinsically efficient
accelerators and that the resulting nonlinear modifications to shock
acceleration (mainly the shock smoothing effect) tend to favour the
acceleration of high rigidity over low rigidity species, that is of
species with higher mass to charge ratios\footnote{This is often
described as a more efficient acceleration of high rigidity
species. However this can be rather misleading. What is meant is that
the {\em steady state} differential velocity spectra of the higher
rigidity species are less rapidly decreasing functions of velocity
than those of the low rigidity species at the crucial low velocities
close to the shock speed. However the low rigidity species have {\em
shorter} acceleration time scales and are accelerated more rapidly,
albeit to steeper spectra.}.  At a crude qualitative level this could
be said to fit the observed enhancement of heavy elements over
hydrogen (e.g., Ellison 1982); however the detailed pattern, and
specifically the atomic physics correlations, are not accounted
for. In addition it is well known that in most of the ISM the
refractory elements are not in the gas-phase but are locked up in the
solid state in interstellar dust grains. From UV absorption line
studies, for example, it is known that the abundance of Fe in the gas
phase is typically only 1\% of its total local ISM abundance. If the
SNR shock is accelerating ions from the ISM gas phase only, how can Fe
be enhanced by a factor of at least 30 relative to hydrogen in the
accelerated particles whereas in the gas phase flowing into the shock
it is generally depleted by a factor of 100?  One could of course
suppose that the bulk of the acceleration occurs in a very hot phase
of the ISM where the grains are destroyed; however there the
characteristic energies are far too high for few eV atomic physics
effects to be important (and, in particular, to select the elements
according to their FIP values); so, in any case, this would merely
undo the grain depletion and not by itself generate an {\em
overabundance} of the refractory/low-FIP elements.

The resolution of this problem, and, we believe, the key to
interpreting the compositional data, is to recognise that dust grains
in the ISM are charged and can therefore be shock accelerated. This is
in fact quite an old idea. Epstein (1980) first suggested that charged
dust grains could be accelerated, and that ions sputtered off the
accelerated grains while the grains were in the upstream region would
then be picked up and further accelerated by the shock\footnote{At
about the same time Cesarsky and Bibring (1981) and Bibring and
Cesarsky (1981) also discussed grain sputtering as a source of GCR
material; however they did not consider grain acceleration and relied
on downstream second order Fermi acceleration to accelerate sputtered
ions.}. With the advances in our understanding of shock acceleration
it is now possible to calculate {\em quantitatively} and in some
detail the process sketched out by Epstein. The full details can be
found in Ellison, Drury and Meyer (1997); here we will concentrate on
conveying the spirit of the calculations and the results.

The essential idea is to apply the modern theory of shock acceleration
consistently to a SNR shock propagating in a dusty ISM. The refractory
elements, such as Iron, Magnesium and Silicon, are known to be almost
entirely condensed into small dust grains with a range of sizes
extending from clusters of a few atoms to a maximum size of about
$10^{-7}\,\rm m$ (this size range is required to fit the UV, optical
and IR data). These grains will be charged by a number of processes
(secondary electron emission, photoelectric effect, plasma charging
etc) to surface potentials of order 10 to $100\,\rm V$ (it is
important to note that this is a standard part of ISM grain theory,
not an assumption of our model; see, e.g. Spitzer 1978) implying mass
to charge ratios for the larger grains of order $10^8$ (and less for
the smaller grains). This means that, relative to a shock moving at
several hundred kilometers per second, their magnetic rigidity is {\em
less} than that of a $10^{14}\,\rm eV$ proton or electron. If the
shock is capable of accelerating particles to these energies, the dust
grains will inevitably also be scattered across the shock and
accelerated.

In fact in the case of at least one remnant, that of the supernova of
1006, there is now direct observational evidence for the acceleration
of electrons to these energies from the detection of X-ray synchrotron
emission (Koyama et al, 1995) and inverse-Compton gamma-rays (Tanimori
et al, 1998). The acceleration of protons to about $10^{14}\,\rm eV$,
although not yet directly observed, is also required if SNRs are to
provide the bulk of the Galactic cosmic ray population up to the
``knee'' energy. Thus it seems certain that at least some supernova
remnant shocks are associated with magnetic field structures capable
of scattering particles of rigidities up to $10^{14}\,\rm V$.  The
dust grains will have the same scattering mean free path as the
ultra-relativistic protons and electrons of the same
rigidity\footnote{ As this is a crucial aspect of the model it is
worth discussing it in a little detail. The key point is that the
trajectory a charged particle follows in a stationary magnetic field
is determined only by the rigidity of the particle. Of course the time
taken to traverse the trajectory will be different for particles of
different velocities, but the path followed is the same for all
particles of the same rigidity (and charge sign). Thus, as long as the
field varies on time scales longer than the time taken by a particle
to traverse the scattering magnetic structures, the scattering mean
free path depends only on the rigidity and will be exactly the same
for a subrelativistic dust grain and a high energy proton. The time
scale for variation in the field will be of order the length scale of
the magnetic structure divided by the Alfv\'en speed, so basically all
particles with velocities larger than the Alfv\'en speed see an
essentially static field and will have scattering mean free paths
which are determined only by their rigidity. The dust grains enter the
downstream region with a velocity of order the shock velocity, and as
the shock is super-Alfv\'enic, this condition is fulfilled for them.} but a
much {\em lower} diffusion coefficient because the grain velocity is only of
order the shock speed and the diffusion coefficient is, within factors
of order unity, just the product of the scattering mean free path and
the particle speed.

The conventional picture is that the magnetic field structures
themselves are generated by plasma instabilities driven by the
accelerated proton pressure gradients in a bootstrap process which
drives the mean free path down to a value of order the gyro-radius
(Bohm scaling). For the subrelativistic dust grains this means that
the effective diffusion coefficient for transport near the shock front
rises as momentum, or velocity, to the second power (one from the
increase in the mean free path and one from that in the speed). Thus as the
grains are accelerated from an initial velocity of order the shock
speed, the acceleration rate drops rapidly. At the same time the
frictional drag on the dust particles resulting from collisions with
atoms of the gas increases proportional to velocity. This sets a
natural limit to the amount of grain acceleration determined by the
balance between acceleration at the shock and frictional losses in the
upstream and downstream regions. This process, which was not
considered by Epstein, turns out to be crucial because it links the
rate of gas collisions with the grain in the upstream region, and
hence the amount of ion sputtering, to the acceleration rate of the
shock. For any reasonable parameters the grains are only slightly
accelerated, by about a factor ten in momentum or velocity, or one
hundred in energy (corresponding to about $0.1\,\rm MeV$ per nucleon),
but this is enough to produce a small amount of sputtering from the
accelerated grains in the region ahead of the shock.  We calculate
that roughly $10^{-4}$ of the grain material will be sputtered in the
upstream region, and give rise to secondary ions with velocities about
ten times the shock speed.  These ions will be picked up by the
interstellar field and swept into the shock, which will then
efficiently accelerate them to relativistic energies.
Particles that are sputtered downstream from the shock are swept away
from the shock and do not get accelerated by it.  An important point is that
the ions sputtered {\em upstream} are produced in association with and close
enough to the
shock to reach it before they have suffered serious energy losses (as
noted above the energy loss times of subrelativistic ions are quite
short).

In addition, of course, the same shock will directly accelerate ions out
of the gas phase, but because these start down in the thermal
distribution at velocities of order the shock speed there will be a
strong rigidity dependent bias in the initial phase of the
acceleration.
Whatever the precise conditions of ionization, this will
effectively result in a mass fractionation effect; in particular if,
as plausible, we have a UV photoionized gas in which all species have
ionization state one or two.
Heavier species have larger mean free paths against scattering and
thus sample more of the shock compression earlier in their
acceleration than the lighter ions. This effect is not easy to model
analytically (x-ref Malkov; Berezhko, Yelshin and Ksenofontov, 1996)
but has been simulated in Don Ellison's Monte-Carlo model of shock
acceleration for many years. Where it has been possible to compare the
Monte Carlo results with observations at the Earth's bow shock the
agreement is generally excellent (eg Ellison, Moebius and Paschmann,
1990).

\begin{figure}
\epsfxsize=\hsize
\epsfbox{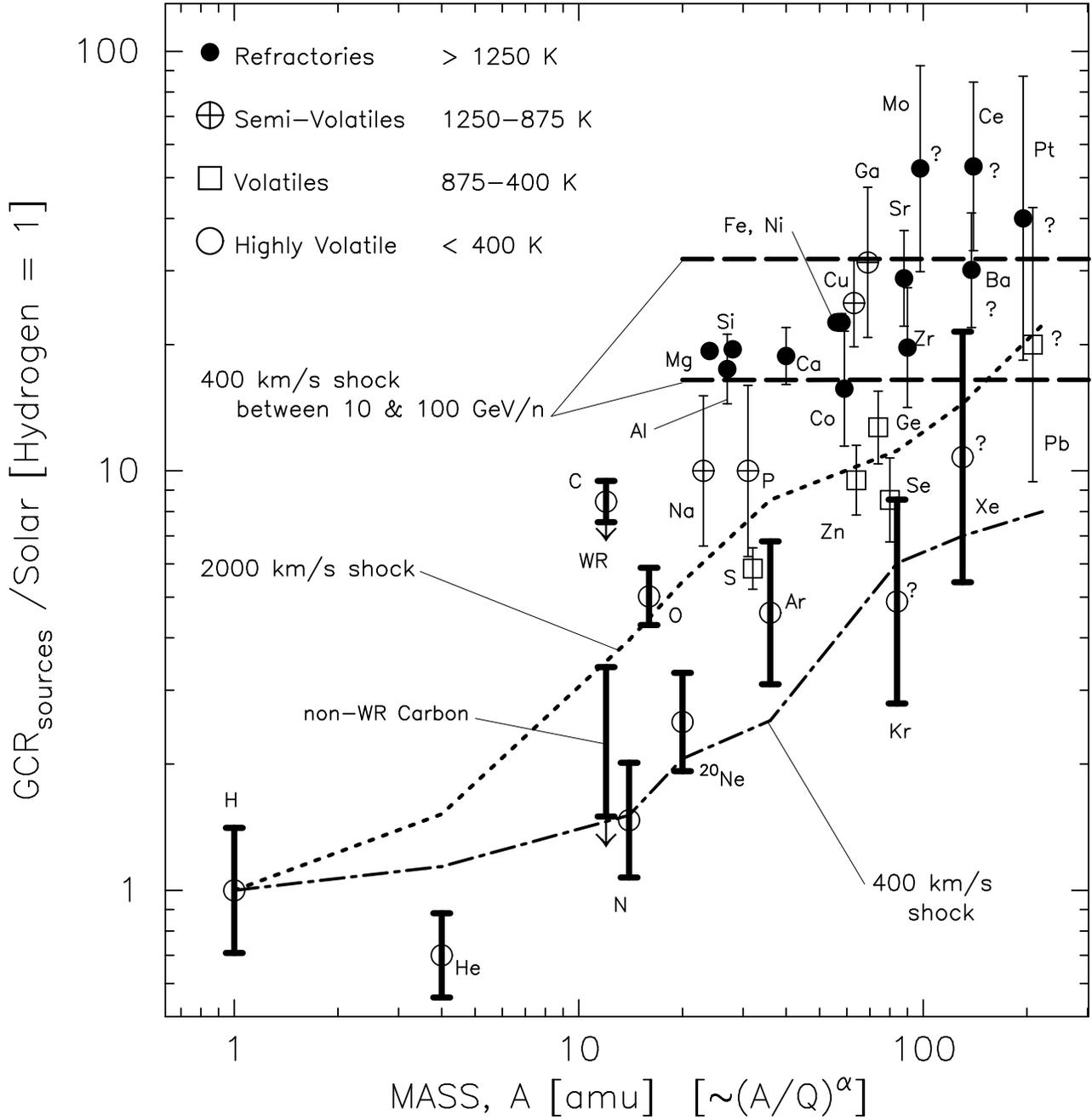}
\caption{\label{ModelPlot} Comparison between the Monte Carlo model
predictions and the observational data. The dotted and dot-dash lines
indicate the mass-dependent fractionation of purely volatile species
given by the code for two different shock speeds.  The two dashed
horizontal lines indicate the enhancement relative to hydrogen of
sputtered ions from refractory grains predicted by the slower shock
model.}
\end{figure}

In terms of their contribution to the bulk composition of GCRs the
most important SNR shocks are thought to be those associated with the
larger older remnants nearing the end of their Sedov-like phase. The
smaller faster shocks associated with young remnants certainly
accelerate particles, but they process relatively small amounts of the
ISM and the particles they accelerate (except at the highest energies)
are trapped inside the remnant and subject to adiabatic losses as the
SNR expands. In fact our results turn out not to be very sensitive to
the assumed shock speed. We have considered two typical cases, a
fast shock of velocity $2000\,\rm km\,s^{-1}$ and a slower older shock
of velocity $400\,\rm km\,s^{-1}$. For both we have calculated the expected
mass fractionation in the acceleration of the volatile element ions
using Ellison's Monte Carlo code and consistently calculated the grain
acceleration using the same code. We then used a simple approximation
for the sputtering yields to estimate the flux of sputtered energetic
ions of refractory elements into the shock, which we then accelerated
to relativistic energies, again using the same Monte Carlo code. The
results are shown in Fig.~\ref{ModelPlot}

This is a very remarkable plot. We see at once that H, the inert gases
and N show a clear mass fractionation effect which is well
reproduced by the Monte Carlo code. In fact the agreement is best with
the slower shock model, exactly as we expect on physical grounds. The
refractory elements, which form the low-FIP group, fall exactly in the
region where we predict accelerated sputtered ions from accelerated
grains to lie (between the two horizontal lines given to roughly
indicate errors in the model).  It is important to note that we have
not done any fitting to the data in this plot. We have simply taken
known physics, a standard dusty ISM composition, and the Monte Carlo
shock acceleration code and calculated {\it ab initio} what the
composition of the accelerated particles should be at GeV per nucleon
energies.

What we have done is to identify {\em two} distinct routes within the
{\em one} shock acceleration mechanism whereby atoms of interstellar
material can reach cosmic ray energies.  The first route is the
standard shock acceleration picture in which the shock-heated
gas-phase ion distribution functions develop tails extending to very
high energies and where high mass to charge ratio ions are
preferentially accelerated.  The second route involves modest
accleration of the interstellar grain population, sputtering of ions
from the accelerated grains upstream of the shock, and subsequent
acceleration of these sputtered ions at higher energies where the
shock acceleration is already quasi-independent of rigidity resulting
in little or no species dependent bias.  The dynamical coupling
between acceleration rate, frictional drag and sputtering links these
two routes and shows that the second route, as measured by abundances
at GeV energies, is about a factor 30 times as efficient as the first
is for protons. This naturally explains the very similar enhancements
of all the low-FIP refractory elements which we believe enter the
cosmic ray population almost exclusively through the second route, and
therefore undergo the crucial first phases of the acceleration, not as
individual ions, but as constituents of entire grains. The volatile
elements, on the other hand, and in particular the dominant species H
and He, enter almost exclusively as gas-phase ions through the first
process and are subject to strong mass fractionation as predicted by
the Monte Carlo model.

\section{Carbon and Oxygen}

That exceptions prove\footnote{In the original sense of {\em test} as
preserved in the saying ``the proof of the pudding is in the eating'}
rules is a principle going back to the scholastic philosophers; one of
the best ways of testing the validity of any general principle is to
look closely at the cases where it apparently fails.  Clearly the two
elements which appear exceptional in our interpretation of the data
are carbon and oxygen. Both are clearly above the ``volatile'' curve,
but also well below the ``refractory'' band. However both these
elements are also special in at least two other respects. Firstly,
although far from being completely condensed in the ISM, both are
known to be important grain constituents. Secondly carbon, and to a
lesser extent oxygen, are greatly enhanced in the winds from
Wolf-Rayet (WR) stars\footnote{Nitrogen is also enhanced in WN-type WR
star winds, but only by factors of order 13. In contrast, carbon is
enhanced by much larger factors of about 120 in the WC-type star
winds. Oxygen can also be enhanced in the much rarer WO-type WR star
winds.}.

Let us consider first the case of oxygen.  The refractory metallic
elements such as Mg, Al, Si, Ca and Fe which form grains do so by
condensing as oxides (silicate minerals are known from infrared
spectroscopy to be one of the main components of interstellar dust).
It is generally estimated that this locks some 15\% to 20\% of the
interstellar oxygen up in grains from which it will be preferentially
accelerated in the same way as all other grain constituents.  Relative
to H, fig .. shows that O is enhanced by a factor of $\fo\approx5$
whereas the neighbouring (in mass) volatile elements N and Ne are only
enhanced by a factor $\fvol\approx2$. However sputtered grain material
is enhanced by a factor $\fref\approx 20$ to 25.  If a fraction $x$ of
the oxygen is in grains, then the resulting mixture of directly
accelerated gas-phase ``volatile'' oxygen and sputtered ``refractory''
oxygen will give a net enhancement \be x \fref + (1- x) \fvol =
\fo\qquad\hbox{so that}\qquad x=(\fo-\fvol)/(\fref-\fvol).  \ee With
the above observed values of the enhancement $f$ factors we get a
fraction $x$ of order 0.15, perfectly consistent with the chemistry of
silicate minerals.  Recent depletion studies of the ISM suggest, if
anything, rather higher fractions of interstellar oxygen in the dust
component (Meyer, Jura and Cardelli, 1998), but this is hard to
reconcile with the chemistry.

The other element which is known to form a significant interstellar
grain component is carbon. Small graphite-like grains have long been a
feature of dust models and are apparenly required to reproduce the
$2175\,$\AA\ feature in the interstellar extinction curve.
Although no generally accepted carrier has yet been identified for the
unidentified interstellar absorption features, almost all suggestions
involve substantial amounts of carbon. Carbon grains are observed to
condense in the atmospheres of carbon stars from which they are
ejected by radiation pressure into the ISM. In
addition, Greenberg and his colleagues have argued that most
interstellar grains acquire ``mantles'' of refractory organic deposits
through condensation of organic compounds in molecular clouds and
subsequent UV irradiation.  In fact at present there is a ``carbon
crisis'' in that the dust models all require more carbon in the dust
than appears to be available (Snow and Witt, 1996; Mathis, 1996; Dwek,
1997). The best current estimates (Cardelli et al, 1996) suggest that
the fraction of ISM carbon incorporated into dust grains is between
20\% and 60\%. The GCRS/solar system enhancement of about 9 relative
to H would be compatible with a fraction of about 30\% of the
interstellar carbon in refractory grains, but in addition we expect a
specific carbon excess from WR star nucleosynthesis.

The existence of such a component is indicated by the the one firm
isotopic anomaly in the GCRS, the well-established excess of
$^{22}$Ne. This strongly suggests a contribution from material
contaminated by the winds from WR stars (van der Hucht and Williams,
1995). This is actually very natural. The most massive supernova
progenitor stars are thought to evolve through a WR phase just prior
to core collapse. The strong and fast WR wind will blow a
circumstellar shell of material enriched in He burning products which
will then be traversed by the subsequent SNR shock wave. This rather
naturally accounts for a $^{22}$Ne excess and must also contribute
significant amounts of $^{12}$C to the GCRS. In fact a large fraction
of the $^{12}$C should condense as grains in the C-rich WR star wind
so that sputtering of these circumstellar grains may further enhance
the importance of this contribution. This may not actually leave much
room for a carbon enhancement from ``ordinary'' ISM grains!
As for oxygen, the WR
contribution of $^{16}$O to the GCRS must be negligible in view of the
observed lack of any associated excess of $^{25,26}$Mg.

\section{Conclusion}

We have shown that standard shock acceleration applied to an ISM with
the bulk composition of the local ISM, but where the refractory
elements along with 15\% of the oxygen and a significant fraction of
the carbon are in dust grains, can replicate {\em all} the observed
features of the GCRS abundance pattern and do so {\em quantitatively},
with the exception of the $^{22}$Ne excess. This latter can be rather
naturally explained in terms of an additional Wolf-Rayet wind
component which must then also contribute to the C excess.

Our interpretation of the compositional data is based solely on
calculable physical processes and standard astronomical inputs with
essentially no adjustable parameters, yet manages to give a better
match to the observations than any other interpretation we are aware
of. This, we feel, argues strongly for the basic correctness of the
underlying picture which locates the origin of bulk of Galactic cosmic
rays in the processing of a dusty ISM by the strong blast waves driven
by supernova explosions.

The exciting prospect is that we appear to be able to relate the GCRS
composition to important astronomical questions about dust and the
ISM.  Obviously much more work needs to be done in refining the model
(in particular the treatment of sputtering needs to be improved) and
this needs to be related to our rapidly increasing knowledge of
interstellar abundances and dust properties. However, we may
ultimately be able to use cosmic ray composition studies to chemically
analyze the interstellar dust!

\section {Acknowledgements}

The work of LD and JPM was supported by the TMR programme of the
European Union under contract FMRX-CT98-0168 and that of DCE by the
NASA Space Physics Theory Program.  LD would like to thank the Service
d'Astrophysique of the CEA/Saclay for their warm hospitality while
working on the first draft of this article. Useful discussions on dust
composition and ISM abundances with Renaud Papoular, Suzanne Madden and
Anthony Jones are gratefully acknowledged.

{\frenchspacing
\parindent=0pt
\parskip= 5 pt plus 5 pt

\vskip 0.3truecm\vskip 0.3truecm\vskip 0.3truecm\vskip 0.3truecm

\centerline{\bf References}

\vskip 0.3truecm\vskip 0.3truecm

Andersson, H. \& Edvardsson, B. 1994 AA 290 590.

Arnett, D 1995, ARA\&A 33 115

Baade, W. \& Zwicky, F. 1934, Proc. Nat. Acad. Sci. U. S. 20 254.

Berezhko, E.~G., Yelshin, V., \& Ksenofontov, L. 1996, Sov. Phys.
JETP, 82, 1,

Bibring, J-P. \& Cesarsky, C. J. 1981, Proc 17th International Cosmic
Ray Conference (Paris) 2 289.

Cesarsky, C. J. \& Bibring, J-P. 1981, IAU Sym 94 ``Origin of Cosmic
Rays'' ed G. Setti, G. Spada \& A. Wolfendale (Dordrecht: Reidel) 361

Edvardsson, B., Andersen, J., Gustafsson, B., Lambert, D. L., Nissen,
P. E. \& Tomkin, J. 1993 AA 275 101.

Ellison, D. C., 1982, Ph.D. Thesis, The Catholic University of
America.

Ellison, D. C., Drury, L. O'C. \& Meyer, J-P. 1997, ApJ 487 197

Ellison, D. C., Moebius, E. \& Paschmann, G. 1990 ApJ 352 376

Epstein, R. I. 1980, MNRAS 193 723

Epstein, R. I. 1981, IAU Sym 94 ``Origin of Cosmic
Rays'' ed G. Setti, G. Spada \& A. Wolfendale (Dordrecht: Reidel) 109

Ginzburg, V. L. \& Syrovatsky, S. I. 1964, ``The origin of cosmic
rays'', english translation by H. S. H. Massey incorporating
revisions by the authors, ed. D. Ter Haar, (Pergamon: Oxford).

van der Hucht, K. A. \& Williams, P. M. eds 1995, IAU Symposium 163
Wolf-Rayet Stars: Binaries, Colliding Winds, Evolution (Dordrecht:
Kluwer)

Koyama, K. et al 1995 Nature 378 255

Meyer, D. M., Jura, M. \& Cardelli, J. A. 1998 ApJ 493 222

Meyer, J-P. 1985, ApJS 57 173.

Meyer, J-P., Drury, L. O'C. \& Ellison, D. C. 1997, ApJ 487 182

Meyer, J-P., Drury, L. O'C. \& Ellison, D. C. 1998, ACE
Workshop, Caltech, Jan.\ 1997, Space Sci.\ Rev., in press.

Spitzer, L. Jr., 1978, ``Physical Processes in the Interstellar Medium'',
(Wiley: New York).

Snow, T. P. \& Witt, A. N. 1996 ApJ 468 L65

Tanimori, T., Hayami, Y., Kamei, S., Dazeley, S. A., Edwards, P. G.,
Gunji, S., Hara, S., Hara, T., Holder, J., Kawachi, A., Kifune, T.,
Kita, R., Konishi, T., Masaike, A., Matsubara, Y., Matsuoka, T.,
Mizumoto, Y., Mori, M., Moriya, M., Muraishi, H., Muraki, Y., Naito,
T., Nishijima, K., Oda, S., Ogio, S., Patterson, J. R., Roberts,
M. D., Rowell, G. P., Sakurazawa, K., Sako, T., Sato, Y., Susukita,
R., Suzuki, A., Suzuki, R., Tamura, T., Thornton, G. J., Yanagita, S.,
Yoshida, T., Yoshikoshi, T. 1998 ApJ 497 L25

Timmes, F. X., Woosley, S. E. \& Weaver, T. A.  1995, ApJS 98 617

Woosley, S. E. \& Weaver, T. A. 1995, ApJS 101 181

} 

\end{document}